# Cosmological test of the Yilmaz theory of gravity

**Michael Ibison**

Institute for Advanced Studies at Austin
4030 West Braker Lane, Suite 300, Austin, TX 78759, USA

Email: ibison@earthtech.org

**Abstract.** We test the Yilmaz theory of gravitation by working out the corresponding Friedmann-type equations generated by assuming the Friedmann-Robertson-Walker cosmological metrics. In the case that space is flat the theory is consistent only with either a completely empty universe, or with a negative energy vacuum that decays to produce a constant density of matter. In both cases the total energy remains zero at all times, and in the latter case the acceleration of the expansion is always negative. To obtain a more flexible and potentially more realistic cosmology the equation of state relating the pressure and energy density of the matter creation process must be different from the vacuum, as for example is the case in the steady-state models of Gold, Bondi, Hoyle and others. The theory does not support the Cosmological Principle for curved space $K \neq 0$ cosmological metrics.

## 1. Introduction

Yilmaz has published two versions of his theory of gravitation. The earlier one introduces an auxiliary scalar field $\varphi$ from which is computed the metric tensor [1], this being the version of the theory cited by Tupper in his short review of scalar-tensor theories [2]. Subsequently Yilmaz published a more comprehensive theory in which the auxiliary field is a tensor [3-8]. In the case of a mass singularity, both theories give an 'exponential metric' with line element

$$ds^2 = \exp(-2\varphi) dt^2 - \exp(2\varphi) d\mathbf{x}^2 \qquad (1)$$

where $\varphi = Gm/r$. This is to be contrasted with the Schwarzschild (GR) line element in isotropic coordinates

$$ds^2 = \left(\frac{1-\varphi/2}{1+\varphi/2}\right)^2 dt^2 - (1+\varphi/2)^4 d\mathbf{x}^2. \tag{2}$$

Comparing the two, one finds for the expansions

$$\begin{aligned}
\text{Yilmaz:} \quad & g_{00} = 1 - 2\varphi + 2\varphi^2 - \frac{4}{3}\varphi^3 \ldots \quad & g_{11} = -1 - 2\varphi - 2\varphi^2 + \ldots \\
\text{GR:} \quad & g_{00} = 1 - 2\varphi + 2\varphi^2 - \frac{3}{2}\varphi^3 \ldots \quad & g_{11} = -1 - 2\varphi - \frac{3}{2}\varphi^2 + \ldots
\end{aligned} \tag{3}$$

To date these give observationally indistinguishable predictions for red-shift, light bending and perihelion advance. Because $g_{00}/g_{xx}$ is never zero, (the coordinate speed of light never vanishes) the Yilmaz metric does not give rise to an event horizon, i.e. the theory does not admit black holes. Unlike GR therefore, there is no limit to the gravitational red-shift of radiation from an arbitrarily dense agglomeration of non-rotating matter. (The maximum possible gravitational redshift from matter in stable orbit of a Schwarzschild hole is $1+z = \sqrt{3/2}$, though arbitrarily large red-shifts are possible from matter in stable orbits of an extreme Kerr hole [9].) Citing this property and assuming the Yilmaz theory is correct, Clapp [10] has suggested that a significant component of quasar red-shift may be gravitational. Robertson [11,12] has suggested that some Neutron stars and Black Hole Candidates may be such 'Yilmaz stars'.

The Yilmaz theory has been extensively investigated by Alley [13,14], who concludes in favour of the theory over GR, in particular because it succeeds in predicting attraction in a simple two-body situation where - allegedly - GR fails to do so. This claim was the subject of a relatively favourable review by Peterson [15]. However, that claim is strongly contested in the paper by Cooperstock and Vollick [16], who maintain that GR is perfectly well-equipped to deal with a two-body problem. The dispute continued with a series of claims and counter-claims involving Fackerell [17-19], and also Misner and Wyss [20-22].

A foundational claim of the Yilmaz theory is that there exists a genuine proper tensor for the gravitational stress-energy. Writing

$$T_\mu^\nu = \tau_\mu^\nu + t_\mu^\nu \tag{4}$$

where $\tau_\mu^\nu$ is the traditional stress-energy tensor of the sources (matter and fields), the additional term $t_\mu^\nu$, which is a function of the metric tensor, stands for the stress-energy of the gravitational field, allegedly improperly omitted in the Einstein theory. Independently, Lo [23] has argued that the formula traditionally used to describe gravitational radiation cannot be derived from Einstein's equations without such a tensor. Though he does not endorse it, Lo cites the Yilmaz theory as a possible example.

A characteristic of the Yilmaz theory is the central role played by an intermediate tensor field $\hat{\varphi}$ which is related to the metric through ([1,3])

$$\mathbf{g} = \mathbf{\eta}\exp\left(2\left(\varphi\hat{\mathbf{1}} - 2\hat{\boldsymbol{\varphi}}\right)\right), \tag{5}$$

where $\mathbf{g} \equiv \{g_{ab}\}$ is the metric tensor, $\mathbf{\eta} \equiv \{\eta_{ab}\}$ is the Minkowski metric, $\hat{\mathbf{1}} \equiv \{\delta_a^b\}$ is the mixed unit tensor, $\hat{\boldsymbol{\varphi}} \equiv \{\varphi_a^b\}$ is a mixed symmetric tensor and $\varphi$ its trace. The field $\hat{\boldsymbol{\varphi}}$ is the solution of the second order equation

$$\Box_B^2 \varphi_a^{\ b} - \frac{1}{\sqrt{-g}}\partial_c\left(\sqrt{-g}\partial^b \varphi_a^{\ c}\right) = 4\pi G \tau_a^{\ b} \tag{6}$$

where $\Box_B^2$ is the Beltrami operator

$$\Box_B^2 = \frac{1}{\sqrt{-g}} \partial_c \sqrt{-g} \partial^c \tag{7}$$

and where $\tau$ is the 'standard' (i.e. commonly used) stress energy tensor of matter, vacuum, and radiation. Writing the Einstein equations as

$$R_a^{\ b} - \frac{1}{2}\delta_a^{\ b} R = -8\pi G \left(\tau_a^{\ b} + t_a^{\ b}\right) \tag{8}$$

and using (6), the Yilmaz decomposition, (4), amounts to the claim that there exists a novel proper tensor $t_a^{\ b}$ satisfying

$$4\pi G t_a^{\ b} = \frac{1}{\sqrt{-g}} \partial_c \left(\sqrt{-g} \partial^b \varphi_a^{\ c}\right) - \Box_B^2 \varphi_a^{\ b} - \frac{1}{2} R_a^{\ b} + \frac{1}{4}\delta_a^{\ b} R \tag{9}$$

where the Ricci tensor and scalar are to be computed using (5). Unfortunately, expressions for $t_a^{\ b}$ supplied by Yilmaz appear to have been derived only after having chosen the harmonic coordinate system, which precludes the possibility of determination of the tensor status of $t_a^{\ b}$ by simple inspection. This appears to be true of (9) because the first two terms on the right hand side (i.e. the left hand side of (6)) are not manifestly covariant.

In their discussion of the existence of the (Yilmaz) stress-energy tensor of the gravitational field, Cooperstock and Vollick draw attention to a requirement by Yilmaz [3] of the matter stress-tensor

$$\partial_\nu \left(\sqrt{-g}\tau^\nu_{\ \mu}\right) = 0 \tag{10}$$

noting that this is not a generally covariant equation, from which they conclude that the theory as a whole is not generally covariant. We observe however, just prior to the statement (10) in [3], Yilmaz makes the particular coordinate choice in favour of the harmonic coordinate system. Consequently one would not expect subsequent relations to be generally covariant, and so it is conceivable that (10) always holds for this coordinate choice in this theory.

Here we will not attempt to adjudicate on the contentious issue of the existence or otherwise of a gravitational stress-energy (true) tensor. Nor will we get involved in the disagreements on what GR has to say about the two-body problem. Instead, given the disagreements above and the consequent uncertain status of the Yilmaz theory, this document is an attempt to provide an independent means of assessing its viability. Herein we report only on what the theory has to say about cosmology and compare its predictions with observation.

## 2. Cosmological metrics

The Cosmological Principle asserts that space should look the same to all co-moving observers. Given that observers have access to (the local values of) the Riemann tensor, the Ricci tensor, and the Ricci scalar, it is inferred that the space parts of these must be expressible in terms of the metric. One finds (see [24] or [25]) that the Riemann tensor must have the form

$$R_{ijkl} = K\left(\gamma_{ik}\gamma_{jl} - \gamma_{il}\gamma_{jk}\right) \tag{11}$$

where $\gamma$ is the 3x3 metric for space and $K$ is a constant. One then has

$$R_{ij} = -2K\gamma_{ij} \tag{12}$$

and

$$R_i^i = 6K. \tag{13}$$

Weinberg [25] shows that any metric having these properties can be written, up to a general coordinate transformation, in the form of the Friedmann Robertson and Walker (FRW) line element which, in the isotropic Cartesian system is

$$ds^2 = dt^2 - \frac{a(t)^2}{\left(1 + K\mathbf{x}^2/4\right)^2} d\mathbf{x}^2, \tag{14}$$

and in 'standard' coordinates ([25]) is

$$ds^2 = dt^2 - a^2(t)\left(\frac{dr^2}{1 - Kr^2} + r^2 d\theta^2 + r^2 \sin^2\theta d\phi^2\right). \tag{15}$$

There is evidence from the WMAP data that space is substantially flat ($\Omega = 1.02 \pm 0.02$ at the $1\sigma$ level [26]), though it is conceded here that the underlying analysis, and therefore the conclusion, may be somewhat model dependent. We first treat this case in detail, postponing consideration of the curved space cosmologies to Section 4. Then one simply has

$$ds^2 = dt^2 - a(t)^2 d\mathbf{x}^2. \tag{16}$$

Generally one sets the time to zero at the present era, and chooses $a(0) = 1$, so that the coordinate speed of light is equal to 1.

Though the Yilmaz theory is allegedly in principle generally covariant, the published details have been worked out only for a harmonic coordinate system, i.e. satisfying

$$\partial_a \left(g^{ab}/\sqrt{-g}\right) = 0 \tag{17}$$

for all *b*, requiring that the metrics corresponding to (16) and (14) be re-cast. A suitable transformation for the flat space FRW case is easily found to be

$$dt = a(\zeta)^3 d\zeta \tag{18}$$

whereupon the resulting metric has line element

$$ds^2 = a(\zeta)^6 d\zeta^2 - a(\zeta)^2 d\mathbf{x}^2 \tag{19}$$

(where, technically, *a* is a new function of its argument – different from that appearing in (16)). With this one then has $\sqrt{-g} = a(\zeta)^6$ from which it is easily seen that (17) is satisfied.

It is important that the cosmological metrics belong to GR no more than they belong to any other metric theory, including that of Yilmaz. In particular it is noteworthy that the author of the Cosmological Principle, Milne, deduced the line element for the hyperbolic expansion without reference to any theory of gravitation [27]. Invoking the Cosmological Principle, Tolman [28] gives a purely deductive derivation of all three cases of expanding metric (positive, negative and zero curvature), and with no reference to Riemann tensors. (See the article by Gale and Urani [29] for an interesting discussion of the historical context.) If one accepts the transcendental status of the Cosmological Principle, then the corresponding geodesic equations of motion in GR - namely the Friedmann equations - serve to identify permissible forms for the *sources*, i.e. the stress-energy tensor. Specifically in the case of GR, in the event that matter is conserved one deduces from the Friedmann equations that the matter in the cosmological fluid is uniformly distributed with constant proper density.

In his early paper Yilmaz [1] proposed a cosmological metric in Cartesian coordinates having line element

$$ds^2 = dt^2 \exp(-\alpha^2 r^2) - \exp(\alpha^2 r^2) d\mathbf{x}^2 . \tag{20}$$

Eq. (20) was adopted and further elaborated on by Mizobuchi [30,31] and in a book by Bjornson [32]. Following Yilmaz [1] there is some discussion in these works attempting to justify a new 'Observation Principle' underpinning this metric. This is at variance, however, with the point of view expressed above, namely the exhaustive and theory-independent status of the FRW metrics. The same conclusion comes from computing the resulting Ricci scalar

$$R = -2\alpha^2 \left(3 + \alpha^2 r^2\right) \exp(-\alpha^2 r^2) ; \tag{21}$$

this metric does not satisfy the Cosmological Principle because all cosmologically co-moving observers do not see the same Riemann tensor, Ricci tensor, and Ricci scalar. The source of the problem can be traced to the specification by Yilmaz in [1] of a matter density that is uniform in the harmonic coordinate system rather than in the proper coordinates. We avoid that error in this document by starting with the cosmological line element (16) and then deducing the form of the stress-energy tensor demanded by the Yilmaz theory.

## 3. Yilmaz Cosmology

### *3.1 Friedmann equations*

Let us first determine the Yilmaz field $\hat{\varphi}$ from the metric tensor by formally inverting Eq. (5) as follows. One has

$$\mathbf{\eta}^{-1}\mathbf{g} = \exp(2\varphi)\exp(-4\hat{\varphi}) . \tag{22}$$

Since the left hand side is a symmetric matrix the standard result relating the determinant to the trace gives

$$\det\left(\mathbf{\eta}^{-1}\mathbf{g}\right) = -g = \exp\left(\mathrm{tr}\left(\log\left(\mathbf{\eta}^{-1}\mathbf{g}\right)\right)\right) = \exp(8\varphi)\exp(-4\,\mathrm{tr}\,\hat{\varphi}) = \exp(4\varphi) . \tag{23}$$

Putting this into (22) gives that the auxiliary Yilmaz tensor field is related to the metric tensor by

$$\exp(-4\hat{\varphi}) = \mathbf{\eta}^{-1}\mathbf{g}/\sqrt{-g} \tag{24}$$

and therefore

$$\hat{\varphi} = \frac{1}{4}\log\left(\sqrt{-g}\,\mathbf{g}^{-1}\mathbf{\eta}\right) . \tag{25}$$

For the cosmological line element (14) this gives that the components of the Yilmaz field are

$$\hat{\varphi} = \log a(\zeta) \, diag(0,1,1,1) . \tag{26}$$

In [3] Yilmaz gives that the energy tensor $\tau$ (here, of matter, vacuum, & radiation) is related to the field $\varphi$ by

$$\Box_B^2 \varphi_a{}^b - \frac{1}{\sqrt{-g}} \partial_c \left( \sqrt{-g} \partial^b \varphi_a{}^c \right) = 4\pi G \tau_a{}^b \tag{27}$$

where $\Box_B^2$ is the Beltrami-Laplace operator. It is asserted in [3] that the second term on the left can in general be set to zero in the manner of a gauge condition [1]:

$$\partial_c \left( \sqrt{-g} \partial^b \varphi_a{}^c \right) = 0. \tag{28}$$

This claim is an error. It turns out (28) is satisfied for the (Schwarzschild) case in the harmonic coordinate system being considered by Yilmaz at the time. But (28) cannot hold in general; there are too many constraints to be satisfied by choice of coordinates. Using (25), we find

$$\sum_c \partial_c \left( \sqrt{-g} \partial^b \varphi_a{}^c \right) = 0 \Rightarrow \partial_\zeta \left( a^6 \partial^\zeta \varphi_\zeta{}^\zeta \right) = 0 \tag{29}$$

(where there are no implied summations in the second expression), so (28) happens to hold in our case. For the Beltrami one has

$$\Box_B^2 \varphi_a{}^b = \sum_c \frac{1}{\sqrt{-g}} \partial_c \left( \sqrt{-g} \partial^c \varphi_a{}^b \right) = \frac{1}{\sqrt{-g}} \partial_\zeta \left( g^{\zeta\zeta} \sqrt{-g} \partial_\zeta \varphi_a{}^b \right)$$
$$= a^{-6} \partial_{\zeta\zeta} \varphi_a{}^b = diag(0,1,1,1) a^{-6} \partial_{\zeta\zeta} \log a(\zeta) \tag{30}$$

Using that the stress-energy tensor for a cosmological fluid is

$$\{\tau_a^b\} = diag(\rho, -p, -p, -p) \tag{31}$$

where $\rho$ is a coordinate density, (27) with (29) and (30) gives the two equations

$$\rho = 0 \tag{32}$$

and

$$\frac{d^2 \log a}{d\zeta^2} = -4\pi G p a^6. \tag{33}$$

Using (18) to return to FRW time, the latter is

$$a^3 \frac{d}{dt}\left( a^3 \frac{d}{dt} \log a \right) = -4\pi G p a^6, \tag{34}$$

which can be simplified with the substitution $b = a^3$ to give

$$\ddot{b} + 12\pi G p b = 0. \tag{35}$$

---

[1] In [5] the relation between $\varphi$ and $\tau$ is given without this term and with no mention of its gauge-like role.

Equations (32) and (34) (or (35)) correspond respectively to the first and second Friedmann equations of GR. We observe that Eq. (10) is automatically satisfied, as claimed by Yilmaz: with no implied summations one has

$$\forall \mu : \sum_\nu \partial_\nu \left(\sqrt{-g}\, \tau^\nu{}_\mu\right) = \partial_\mu \left(a^6 \tau^\mu{}_\mu\right) = \delta^\zeta_\mu \partial_\zeta \left(b^2 \rho\right) = 0, \qquad (36)$$

the last step following from (32).

In the context of GR the second Friedmann equation is a statement of energy conservation, and determines the dependency on the scale factor of the various contributions (i.e. from matter, radiation, and vacuum) to the energy density, assuming that, at least since the radiation era, there is negligible coupling between them, and given an equation of state for each. Under these conditions in GR one thereby obtains ([33],[34])

$$d\left(a^3 \rho_i\right) + p_i\, da^3 = 0. \qquad (37)$$

From this and the equation of state $p_i = k_i \rho_i$, where $i \in \{m, r, v\}$ and $k_i \in \{0, 1/3, -1\}$ corresponding to matter, radiation, and vacuum respectively, one then infers the (usual) scaling of matter radiation and vacuum:

$$\rho_m = \rho_m(0)/a^3(t), \quad \rho_r = \rho_r(0)/a^4(t), \quad \rho_v = \rho_v(0). \qquad (38)$$

By contrast, the Yilmaz Friedmann equations, do not furnish a direct relation between the pressure and energy of the different species in the manner of (37), and therefore one is unable to determine scaling laws of the kind (38) in a straightforward manner, but must instead proceed along different lines.

### 3.2 Empty universe

Decomposing the stress-energy tensor into contributions from matter, radiation, and vacuum, the first Friedmann equation for the Yilmaz theory is

$$\rho = \rho_m + \rho_r + \rho_v = 0. \qquad (39)$$

Matter and radiation have positive energy densities. If one maintains the usual (GR) view that the vacuum energy, if it exists (i.e. plays a cosmological role), is positive, then $\rho_m \geq 0$, $\rho_r \geq 0$, $\rho_v \geq 0$, from which it must be concluded that the Yilmaz theory permits only $\rho_m = \rho_r = \rho_v = 0$, i.e. an empty universe. The equations of state then give that the pressures are also zero. In that case the solution of (35) is

$$b = b_0 + b_1 t \qquad (40)$$

and recalling $b = a^3$ one then has

$$a = \left(1 + 3H_0 t\right)^{1/3}. \qquad (41)$$

### 3.3 Negative energy vacuum

It can be argued that the total vacuum energy is of uncertain sign because there are infinite contributions from the zero-points fields of Fermions and Bosons with opposite signs, whose magnitudes are unrelated. From this perspective the relatively recent position that the vacuum energy is positive rather than zero, say, can be attributed to the requirement that the first Friedmann equation of GR accord with the observed acceleration in expansion. That is, the vacuum has been fitted *a posteriori* to the data, given the theory (GR). Granting the same license to the Yilmaz theory, and admitting the existence of matter and radiation, one would infer from (39) that the vacuum energy is negative. Further, if quantum field theory presents

*independent* reasons for demanding that the energy density of the ZPF is constant during expansion, then (39) demands that the sum of two quite different quantities, $\rho_m + \rho_r$ is a constant:

$$\rho_v = -\rho_m - \rho_r = \text{constant}. \tag{42}$$

Once matter and radiation have decoupled, barring some coincidence (42) implies that $\rho_m$ and $\rho_r$ must individually be constant. Using the equations of state (35) can be written

$$\ddot{b} = -4\pi G b (3\rho_m + 4\rho_r). \tag{43}$$

The solutions are sinusoids with radian frequency

$$\omega = \sqrt{4\pi G (3\rho_m + 4\rho_r)}, \tag{44}$$

and therefore, disregarding an arbitrary phase,

$$a = a_{\max} \sin^{1/3}(\omega t). \tag{45}$$

It is concluded that the theory predicts an oscillating universe and deceleration for all time except perhaps at the singular points at $a = 0$. The deceleration parameter is

$$q \equiv -\frac{\ddot{a}a}{\dot{a}^2} = \frac{3}{\cos^2(\omega t + \phi)} - 1. \tag{46}$$

Using that the Hubble parameter is

$$H = \frac{\dot{a}}{a} = -\frac{\omega}{3}\cot(\omega t + \phi) \tag{47}$$

then at the present time one has

$$q_0 = 3\left[1 + \left(\frac{\omega}{3H_0}\right)^2\right] - 1 = 2 + \frac{4\pi G}{3H_0^2}(3\rho_{m0} + 4\rho_{r0}) = 2 + \frac{3}{2}\Omega_{m0} + 2\Omega_{r0} \tag{48}$$

where

$$\Omega_{i0} \equiv \frac{8\pi G}{3H_0^2}\rho_{i0} \tag{49}$$

and $\Omega = 1$ is the critical density as defined by its role in the (first) Friedmann equation in GR. The SN 1a data permit estimates of the deceleration parameter that are relatively independent of the cosmological model and consistently indicate a negative value of $q_0$, [35-37], at variance with the prediction above that $q \geq 2$ for all time.

*3.4 Explicit matter creation*
Despite the absence of an explicit mechanism, the interpretation of the negative energy vacuum solution must be that the vacuum is decaying to produce matter and radiation in such a way as to maintain a constant coordinate density throughout the expansion. That is, the vacuum is driving a steady-state cosmology. By contrast, an explicit mechanism for the creation of matter and (perhaps) radiation presumably entails introduction of a novel contribution to the stress-energy tensor, having an associated *negative* energy density $\rho_c$ (where $\rho_c < 0$) and a pressure $p_c$, say. Hoyle et al [38] for example use $p_c = \rho_c < 0$, though there seems to be no a priori theoretical constraint on the relation between these two.

Given this, the unknown contribution from the vacuum can be subsumed into these two whereupon (42) becomes

$$-\rho_c = \rho_m + \rho_r. \tag{50}$$

Using the equation of state for the creation terms $p_c = k_c \rho_c$ the total pressure is

$$p = p_m + p_r + p_c = \rho_r/3 + k_c \rho_c = (1/3 - k_c)\rho_r - k_c \rho_m. \tag{51}$$

Therefore (35) is

$$\ddot{b} + 4\pi G\left((1-3k_c)\rho_r - 3k_c \rho_m\right)b = 0. \tag{52}$$

The evolution of the scale factor determined from (52) is quite arbitrary unless, for example, it is known how the matter creation process depends on the local matter density.

### 3.5 Steady-state
In the original 'classical' steady-state cosmology the coordinate densities of matter and radiation are constant. In this case one has that $b$ is harmonic or hyperbolic according to the sign of the total pressure. In the event that the total pressure is negative one has

$$b = b_+ e^{\alpha t} + b_- e^{-\alpha t} \tag{53}$$

where

$$\alpha = \sqrt{4\pi G\left(3k_c \rho_m + (3k_c - 1)\rho_r\right)} \tag{54}$$

and $b_\pm$ are constants. Recalling $b = a^3$, one obtains

$$a = \left(\frac{b_+ e^{\alpha t} + b_- e^{-\alpha t}}{b_+ + b_-}\right)^{1/3}. \tag{55}$$

Setting $a(0) = 1$ and $\dot{a}(0)/a(0) = H_0$ one obtains

$$a = \left(\cosh(\alpha t) + \frac{3H_0}{\alpha}\sinh(\alpha t)\right)^{1/3}. \tag{56}$$

Unlike GR, given constant matter and radiation densities the Yilmaz theory does not automatically generate an exponential expansion, and therefore the resulting cosmology does not automatically satisfy the Perfect Cosmological Principle. Noting that

$$3H_0/\alpha = \sqrt{\frac{9H_0^2}{4\pi G\left(3k_c \rho_m + (3k_c - 1)\rho_r\right)}} = \sqrt{\frac{2}{k_c \Omega_{m0} + (k_c - 1/3)\Omega_{r0}}}, \tag{57}$$

the condition that the growth is purely exponential, $3H_0/\alpha = 1$, is that

$$k_c \Omega_{m0} + (k_c - 1/3)\Omega_{r0} = 2, \tag{58}$$

in which case $a = \exp(\alpha t/3)$ and $q = -1$. In the absence of a physical theory describing the matter creation process there are no constraints on $k_c$ and so (58) can be satisfied for any observed matter density. Given the observed fact that $\Omega_{r0} \ll \Omega_{m0}$ one deduces that $k_c$ is positive. Implementation of the classical steady state cosmology by the Yilmaz theory therefore requires that both the energy density and pressure associated with the creation stress-energy are negative.

## 4. Curved space

### 4.1 Harmonic coordinates

Starting with the FRW line element in standard coordinates (15) we seek new coordinates $\bar{x}^\mu = (\bar{t}, \bar{r}, \bar{\theta}, \bar{\phi})$, each potentially a function of $(t, r, \theta, \phi)$, such that

$$\frac{\partial}{\partial \bar{x}^a}\left(\sqrt{-\bar{g}}\,\bar{g}^{ab}\right) = 0 \tag{59}$$

where $\bar{g}$ is the metric after the coordinate transformation. Since $g$ is diagonal, it is reasonable to attempt a transformation satisfying (59) wherein $\bar{g}$ is diagonal. Assuming such a transformation exists (59) becomes

$$\forall a: \frac{\partial}{\partial \bar{x}^a}\left(\sqrt{-\bar{g}}\,/\,\bar{g}_{aa}\right) = 0, \tag{60}$$

where no sum is now implied. Given that $g$ depends only on $t$, $r$ and $\theta$, the following dependencies are reasonable

$$\bar{t} = T(t),\ \bar{r} = R(r),\ \bar{\theta} = \Theta(\theta),\ \bar{\phi} = \phi. \tag{61}$$

With these the line element (15) becomes

$$ds^2 = \frac{dT^2}{\dot{T}^2} - a^2(t)\left(\frac{dR^2}{\dot{R}^2(1-Kr^2)} + \frac{r^2 d\Theta^2}{\dot{\Theta}^2} + r^2 \sin^2\theta\, d\phi^2\right). \tag{62}$$

The new metric tensor has

$$\sqrt{-\bar{g}} = \frac{a^3 r^2 \sin\theta}{\dot{T}\dot{R}\dot{\Theta}\sqrt{1-Kr^2}}, \tag{63}$$

and therefore the quantity appearing in (60) is

$$\sqrt{-\bar{g}}/\bar{g}_{aa} = \mathrm{diag}\left(\frac{a^3 r^2 \dot{T}\sin\theta}{\dot{R}\dot{\Theta}\sqrt{1-Kr^2}},\ -\frac{ar^2 \dot{R}\sqrt{1-Kr^2}\sin\theta}{\dot{T}\dot{\Theta}},\ -\frac{a\dot{\Theta}\sin\theta}{\dot{T}\dot{R}\sqrt{1-Kr^2}},\ -\frac{a}{\dot{T}\dot{R}\dot{\Theta}\sqrt{1-Kr^2}}\right). \tag{64}$$

Satisfaction of (60) therefore requires

$$\dot{T}a^3 = c_0,\quad r^2 \dot{R}\sqrt{1-Kr^2} = c_1,\quad \dot{\Theta}\sin\theta = c_2 \tag{65}$$

where the $c_i$ are constants. We will choose $c_0 = 1$, $c_1 = -1/|K|$, $c_2 = 1$ (which choice is valid only if $K \neq 0$). The new time coordinate is simply

$$T = \int dt/a^3. \tag{66}$$

The new radial coordinate is

$$R = -\int dr \frac{1}{|K|r^2\sqrt{1-Kr^2}} = \frac{\sqrt{1-Kr^2}}{|K|r} \tag{67}$$

and it will be useful to note that

$$r^2 = \frac{1}{K(1+KR^2)}, \quad \dot{R}^2(1-Kr^2) = K^2 r^4 = \frac{1}{(1+KR^2)^2}. \tag{68}$$

Note the mapping of domains implied by (67):

$$\begin{aligned} K > 0 : r \in \left[0, 1/\sqrt{K}\right] \mapsto R \in [\infty, 0] \\ K < 0 : r \in [0, \infty] \mapsto R \in \left[\infty, 1/\sqrt{-K}\right] \end{aligned}. \tag{69}$$

The new azimuth angle is

$$\Theta = \int d\theta \frac{1}{\sin\theta} = \log(\tan\theta/2) \tag{70}$$

where the domains are

$$\theta \in [0, \pi] \mapsto \Theta \in [-\infty, \infty]. \tag{71}$$

It will be useful to note that

$$\frac{1}{\dot{\Theta}^2} = \sin^2\theta = \left(\frac{2\tan\theta/2}{1+\tan^2\theta/2}\right)^2 = \frac{4e^{2\Theta}}{\left(1+e^{2\Theta}\right)^2} = \frac{1}{\cosh^2\Theta}. \tag{72}$$

Putting Eqs. (66)-(72) in the transformed line element (62) gives that the FRW line element can be expressed in harmonic coordinates as

$$ds^2 = a^6 dT^2 - \frac{a^2}{(1+KR^2)}\left(\frac{dR^2}{(1+KR^2)} + \frac{d\Theta^2 + d\phi^2}{K\cosh^2\Theta}\right). \tag{73}$$

Note that, due to (69), the signs of the contributions remains as $(+ - - -)$ for either sign of $K$.

*4.2 Friedmann equations*
The new line element (73) gives

$$\sqrt{-\bar{g}} = \frac{a^6}{|K|(1+KR^2)^2 \cosh^2\Theta}, \tag{74}$$

and therefore the quantity appearing both in (60) and (64) is

$$\sqrt{-\overline{g}}/\overline{g}_{aa} = diag\left(\frac{1}{|K|(1+KR^2)^2\cosh^2\Theta}, -\frac{a^4}{|K|\cosh^2\Theta}, -\frac{\text{sgn}(K)a^4}{1+KR^2}, -\frac{\text{sgn}(K)a^4}{1+KR^2}\right). \tag{75}$$

Inserting this into (25) gives the Yilmaz tensor field

$$\hat{\varphi} = \frac{1}{4}diag\left(\log\left(\frac{1}{|K|(1+KR^2)^2\cosh^2\Theta}\right), \log\left(\frac{a^4}{|K|\cosh^2\Theta}\right), \log\left(\frac{\text{sgn}(K)a^4}{1+KR^2}\right), \log\left(\frac{\text{sgn}(K)a^4}{1+KR^2}\right)\right). \tag{76}$$

In the case that $g$ and $\hat{\varphi}$ are diagonal and the coordinates are harmonic, the 'gauge part' of (27) is

$$\frac{1}{\sqrt{-\overline{g}}}\sum_c \partial_c\left(\sqrt{-\overline{g}}\partial^b\varphi_a{}^c\right) = \frac{1}{\sqrt{-\overline{g}}}\partial_a\left(\sqrt{-\overline{g}}\,\overline{g}^{bb}\partial_b\varphi_a{}^a\right). \tag{77}$$

An urgent problem is that this expression now has off-diagonal elements and is not symmetric. Given that the Beltrami is diagonal it is concluded that the theory demands that the stress energy tensor also lack symmetry, in conflict with the cosmological requirement of homogeneity and isotropy. This means that all observers do not see the same distribution of energy, even though such a distribution is called for in order to sustain a curvature tensor that *is* the same for all observers. One must conclude that the Yilmaz theory does not support the curved space implementation of the Cosmological Principle.

Given the above, one might wonder if Yilmaz was correct to drop the problematic 'gauge part' in, for example [5], because it should never have been present. If so, would the theory then admit the curved space cosmological metrics? To answer this question, we first observe that (27) would then simply be

$$\Box_B^2 \varphi_a{}^b = 4\pi G \tau_a{}^b . \tag{78}$$

Given that $\hat{\varphi}$ is diagonal the Beltrami can be written

$$\Box_B^2 \varphi_a{}^b = \sum_c \frac{1}{\sqrt{-\overline{g}}}\partial_c\left(\sqrt{-\overline{g}}\partial^c\varphi_a{}^b\right) = \delta_a^b\frac{1}{\sqrt{-\overline{g}}}\sum_c \partial_c\left(\frac{\sqrt{-\overline{g}}}{\overline{g}_{cc}}\partial_c\varphi_a{}^a\right) = \delta_a^b\sum_c \overline{g}^{cc}\partial_c^2\varphi_a{}^a, \tag{79}$$

the last step following because the coordinates are harmonic. It will be sufficient to examine just the $(T,T)$ component of this expression:

$$\Box_B^2\varphi_T{}^T = \sum_c \overline{g}^{cc}\partial_c^2 \log\left(\frac{1}{|K|(1+KR^2)^2\cosh^2\Theta}\right)$$

$$= -\frac{1}{4a^2}\left((1+KR^2)^2\partial_R^2 + (1+KR^2)K\cosh^2\Theta\,\partial_\Theta^2\right)\log\left(\frac{1}{|K|(1+KR^2)^2\cosh^2\Theta}\right). \tag{80}$$

$$= \frac{K}{2a^2}\left(2(1+KR^2) - 4KR^2 + (1+KR^2)\cosh^2\Theta\left(1-\frac{\sinh^2\Theta}{\cosh^2\Theta}\right)\right)$$

$$= \frac{K}{2a^2}(3-KR^2)$$

Inserting this into (78) gives

$$K(3 - KR^2) = 8\pi G \rho a^2. \tag{81}$$

Once again, this is a problem: according to the Yilmaz theory a cosmological ($K \neq 0$) metric that is allegedly the same for all observers must be sustained by non-uniform distribution of energy. The conclusion, therefore, is the same as that obtained with the gauge part retained: the Yilmaz theory does not support the curved space implementation of the Cosmological Principle.

## 5. Discussion and conclusions

### 5.1 Flat-space metric
For a flat space FRW metric the Yilmaz theory demands that the total energy density (vacuum, matter, radiation, etc) is zero. In a matter conserving cosmology, if the vacuum energy density is assumed positive, the Yilmaz theory is compatible only with a universe where the total energy density is devoid of matter, radiation, and vacuum energy. Alternatively, if the vacuum energy density is allowed to take either sign, the Yilmaz theory predicts a cosmology that is always decelerating, with $q \geq 2$. Both of these are at variance with observation.

Whereas in the latter case the theory requires that a (negative energy) vacuum supply the matter and radiation required to keep their densities constant, a steady-state theory admitting an explicit matter creation stress-energy term having negative energy density but with arbitrary pressure permits an arbitrary evolution of the scale factor.

### 5.2 Curved-space metrics
The Yilmaz theory requires a non-uniform distribution of matter in order to generate the curved-space FRW metrics. That is, whilst all co-moving observers everywhere see the same evolution of the scale factor, they do not see the same local distribution of matter. Since both curvature and matter distribution are components of the local environment, this means that all co-moving observers do not see the same universe. It must be concluded that the Yilmaz theory does not accommodate the curved space instantiations of the Cosmological Principle.

### 5.3 The Yilmaz gravitational stress-energy tensor
This paper did not address directly the existence or otherwise of a true gravitational stress-energy tensor as claimed by Yilmaz. If such a gravitational stress-energy tensor does exist, its weight, relative to the Einstein tensor, is initially a free parameter. In [5] Yilmaz argues at some length for a particular weight, which is associated therein with $\lambda = 1$, $\lambda$ being the free parameter, and $\lambda = 0$ corresponding to Einstein GR. Given the predictive observational successes of GR it is hardly likely that there can be any decisive observational arguments in favour any particular value of $\lambda$ other than 0 – at least outside of cosmology. It follows that other values of $\lambda$ should, a priori, be regarded on an equal footing with Yilmaz' choice of $\lambda = 1$. Although the mathematical details of this possibility are not developed in the published work, nonetheless one can deduce that for intermediate (i.e. $0 < \lambda < 1$) values of $\lambda$, a hybrid Yilmaz-Einstein theory will no longer demand that the total mass-energy density vanish, and therefore such a theory could possibly generate a viable cosmology. Further, given the agreement of both Einstein ($\lambda = 0$) and Yilmaz ($\lambda = 1$) theories with observations of light bending and perihelion advance, and barring a strange mathematical coincidence, the hybrid theory will generate the same agreement throughout the intermediate domain.

**Acknowledgements**

The author is grateful to Harold Puthoff for some very useful discussions on the issues in this paper.